# Monolithically integrated photonic neural network at driven-dissipative criticality

Yuming Zhang[1]†, Ruoran Wang[1]†, Jingcheng Li[1], Hailong Zhou[1,2*], Jianji Dong[1,2*], Zhipei Sun[3*], Xinliang Zhang[1]

[1]*Wuhan National Laboratory for Optoelectronics, Huazhong University of Science and Technology, Wuhan 430074, China*

[2]*Optics Valley Laboratory, 430074 Wuhan, China*

[3]*Department of Electronics and Nanoengineering, Aalto University, Espoo, Finland*

[*]Corresponding author: hailongzhou@mail.hust.edu.cn, jjdong@mail.hust.edu.cn, zhipei.sun@aalto.fi

## Abstract

Artificial intelligence increasingly demands computing architectures capable of adaptive representation, temporal information processing, and robust operation under uncertainty. However, conventional photonic neural architectures typically rely on predefined computational operations and separate functional modules, limiting the ability of physical systems to exploit their intrinsic dynamics for intelligence. Here, we demonstrate a monolithically integrated active photonic neural network (APNN) that harnesses driven-dissipative criticality as a physical computational resource. By balancing optical excitation and dissipation, the system operates near a critical regime where weak input-dependent variations are amplified into separable representations while stable attractor dynamics are preserved. The APNN integrates an optoelectronic nonlinear array and a reconfigurable symmetric optical coupling matrix on a silicon photonic chip, enabling recurrent physical computation within a compact closed loop. We demonstrate nonlinear classification with 95.2% accuracy using only 100 training samples and 44 trainable parameters. And it remains robust at an ultralow signal-to-noise ratio of -30 dB. Beyond static recognition, the same driven-dissipative dynamics enable long-range temporal feature extraction from electrocardiogram signals and associative recovery of corrupted patterns through attractor-based relaxation. These results establish driven-dissipative criticality as a unified mechanism for representation enhancement, temporal information retention, and error-resilient computation, providing a pathway toward scalable photonic systems in which intelligence emerges from intrinsic physical dynamics.

## Introduction

Artificial intelligence increasingly requires computing systems that distinguish complex signals, retain information over time, and operate reliably under noise [1]. Physical neural networks compute through the intrinsic dynamics of a substrate [2-7], but face a central trade-off between dynamical sensitivity and stability. Greater sensitivity amplifies small input differences and enriches representations, whereas excessive sensitivity can destabilize evolution; strong stabilization, in turn, can suppress state diversity and memory.

Driven-dissipative systems provide a way to navigate this trade-off. The balance between external energy injection and intrinsic dissipation controls their operating regime. Near criticality, responses to small input differences can be enhanced [8-10], while stability also depends on recurrent coupling and nonlinear response. Although critical phenomena have been widely studied, how to exploit them for computation while preserving stable relaxation within an integrated physical system remains an open question.

Here, we demonstrate a monolithically integrated active photonic neural network (APNN) that exploits driven-dissipative criticality for physical computation. The system integrates an optoelectronic nonlinear array (OENA) and an optical symmetric coupling matrix (OSCM) on a silicon photonic chip to form a closed recurrent loop. The OENA provides nonlinear state evolution and an actively tunable optical drive, whereas the OSCM implements symmetric recurrent coupling that structures the dynamical landscape. By operating near the critical boundary, the APNN simultaneously achieves enhanced input sensitivity, extended temporal information retention, and attractor-based state recovery. We experimentally demonstrate nonlinear classification with 95.2% accuracy using only 100 training samples and 44 trainable parameters, temporal feature extraction from electrocardiogram signals, and associative recovery of corrupted patterns. These results establish driven-dissipative criticality as a unified physical mechanism for expressive, robust, and hardware-efficient photonic intelligence.

## Principle

The APNN connects an OENA and an OSCM in a closed loop (Fig. 1A). We implemented the APNN on a 220 nm silicon-on-insulator (SOI) platform, integrating an 8 × 8 OSCM, an 8 × 1 OENA, and optical feedback waveguides on a single chip (Fig. 1F). At each round trip, the OENA applies a nonlinear response to the input and current state, whereas the OSCM mixes the node responses through a trainable symmetric matrix. An external drive offsets loop dissipation and controls the feedback strength.

Taking one loop round trip as a discrete time step, the state dynamics are modeled by:

$$\boldsymbol{s}^{(\boldsymbol{t}+\mathbf{1})} = (1-\lambda)\boldsymbol{s}^{(\boldsymbol{t})} + \lambda\eta_{out}G(P_{pump})\boldsymbol{W}f\big(\eta_{in}\boldsymbol{s}^{(\boldsymbol{t})} + \boldsymbol{x}_{in} + \boldsymbol{b}\big) + \boldsymbol{n}^{(t)} \tag{1}$$

where $\boldsymbol{s}^{(t)}$ is the network state after roundtrip $t$, $G(P_{pump})$ is the active gain controlled by the external pump power, $\boldsymbol{W}$ is the symmetric coupling matrix, $f(\cdot)$ is the nonlinear transfer function, $\boldsymbol{x}_{in}$ is the input feature, $\boldsymbol{b}$ is the static bias setting the OENA operating point, $\eta_{in}$ and $\eta_{out}$ represent the effective input coupling and feedback efficiencies determined by optical transmission losses and device responses, and $\boldsymbol{n}^{(t)}$ is the additive perturbation. The effective update rate $\lambda \in (0,1]$ characterizes the bandwidth-limited closed-loop response of the OENA, while $\eta_{in}$ and $\eta_{out}$ represent the effective input coupling and feedback efficiencies determined by optical transmission losses and device responses. The first term in Eq.

1 retains the previous state, whereas the second introduces new feedback from nonlinear transformation and recurrent coupling. The update factor $\lambda \in (0,1]$ is determined by the device response time and the loop round-trip time. A smaller λ corresponds to slower state adjustment per round trip.

We define the maximum amplification factor per roundtrip as the local effective feedback gain $g_{loop}$, which is jointly determined by the external pump, distributed loss, symmetric coupling and the local nonlinear slope of the OENA. Near the critical operating regime of $g_{loop} \approx 1$, weak input-dependent differences among initially compact states are progressively amplified and expanded into more discriminative representations, which subsequently evolve towards distinct stable attracting regions (Fig. 1B). Meanwhile, the influence of an input $\boldsymbol{x}_t$ can propagate through successive recurrent state updates, and the finite device bandwidth causes this influence to decay gradually over multiple evolution steps, while active pumping compensates recurrent loss and extends its effective retention time (Fig. 1C). Under stable operating conditions, clean and perturbed states within the same attraction basin can relax towards the same stable attractor, yielding consistent outputs and perturbation-tolerant readout (Fig. 1D). Representation expansion, temporal information retention and stable recovery are therefore jointly supported by the same driven-dissipative recurrent dynamics, providing the physical basis for the classification, temporal-processing and associative-memory tasks examined below.

To quantify these mechanisms, we simulated the network response at different pump strengths (Fig. 1E). Near a steady state, small state deviations evolve as $\delta \boldsymbol{s}^{(\boldsymbol{t}+\boldsymbol{1})} = \boldsymbol{A}\delta \boldsymbol{s}^{(\boldsymbol{t})}$, where $\boldsymbol{A}$ is the local state-update matrix. Its spectral radius $\rho(\boldsymbol{A})$ is the largest absolute value of its eigenvalues. When $\rho(\boldsymbol{A}) < 1$, small state deviations caused by a transient noise perturbation decay, and the steady state is locally stable. We quantify the steady-state response using the input sensitivity $\chi_s = \left\| \frac{\partial \boldsymbol{s}^*}{\partial \boldsymbol{x}} \right\|_2$. A larger $\chi_s$ indicates a stronger steady-state response to small input differences. We use the pulse retention time to measure how long the influence of a brief input persists after the input ends. As the pump strengthens and the local effective feedback gain $g_{loop}$ approaches unity from below, both sensitivity and pulse retention time increase and peak in the near-critical regime, while $\rho(\boldsymbol{A})$ remains below unity. At higher pump powers, nonlinear saturation weakens the effective feedback, causing both quantities to decline. These results show that pump control can enhance input responses and extend information retention while preserving local stability, beyond simply compensating optical loss.

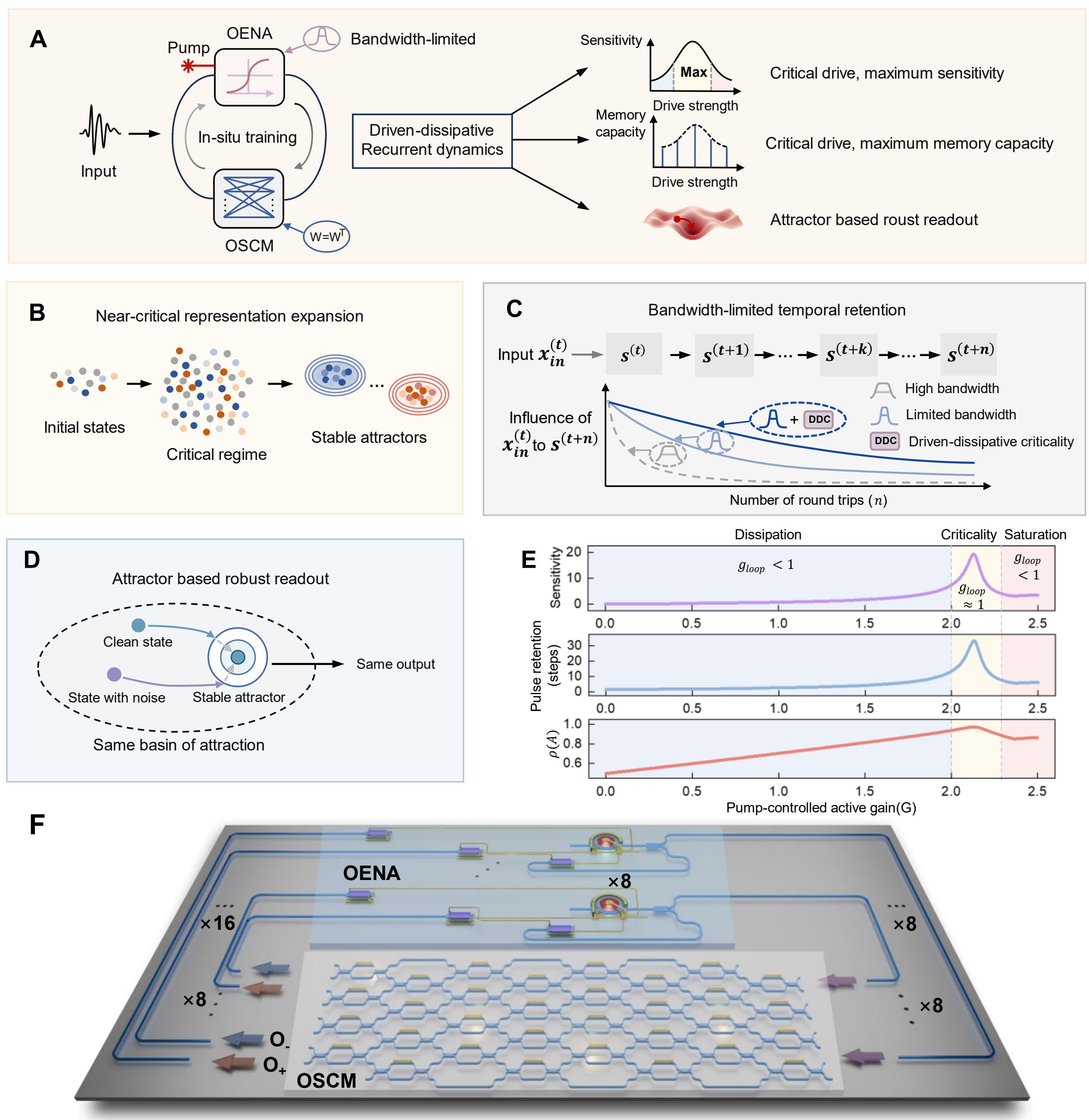

**Fig. 1. Driven-dissipative principle of the APNN.** (**A**) Conceptual architecture comprising an optoelectronic nonlinear array (OENA) and an optical symmetric coupling matrix (OSCM). $\boldsymbol{W} = \boldsymbol{W}^T$ denotes symmetric recurrent coupling. (**B**) Schematic evolution of input-dependent states. Points represent network states, and contours indicate distinct attracting regions. (**C**) Schematic persistence of an input's influence on subsequent network states. Gray dashed, light blue, and dark blue curves represent high-bandwidth operation, bandwidth-limited operation, and bandwidth-limited operation with driven-dissipative criticality, respectively. n denotes the number of round trips since the input was applied. (**D**) Schematic recovery within a common basin of attraction, outlined by the dashed boundary. Blue and purple arrows trace the evolution of unperturbed and perturbed states toward the same stable attractor. (**E**) Representative model simulations as a function of in-loop active gain $G$. From top to bottom: steady-state input sensitivity, pulse retention time, and the spectral radius $\rho(\boldsymbol{A})$ of the full state-update matrix. Blue, pale yellow, and pink shading indicate dissipation-dominated, near-critical, and saturation-dominated regimes. Pulse retention time is defined as the number of update steps required for the pulse-response norm to decay from its peak to 1/e of that value. (**F**) Chip schematic showing the 8 × 8 OSCM, eight OENA units, and feedback waveguides. Arrows indicate propagation directions; multiplicity labels give channel or unit counts.

## Results

To examine the computational scope of the APNN, we selected three tasks with distinct input structures and readout requirements (Fig. 2). These tasks evaluate nonlinear discrimination, the use of retained temporal information, and the recovery of target patterns from corrupted cues.

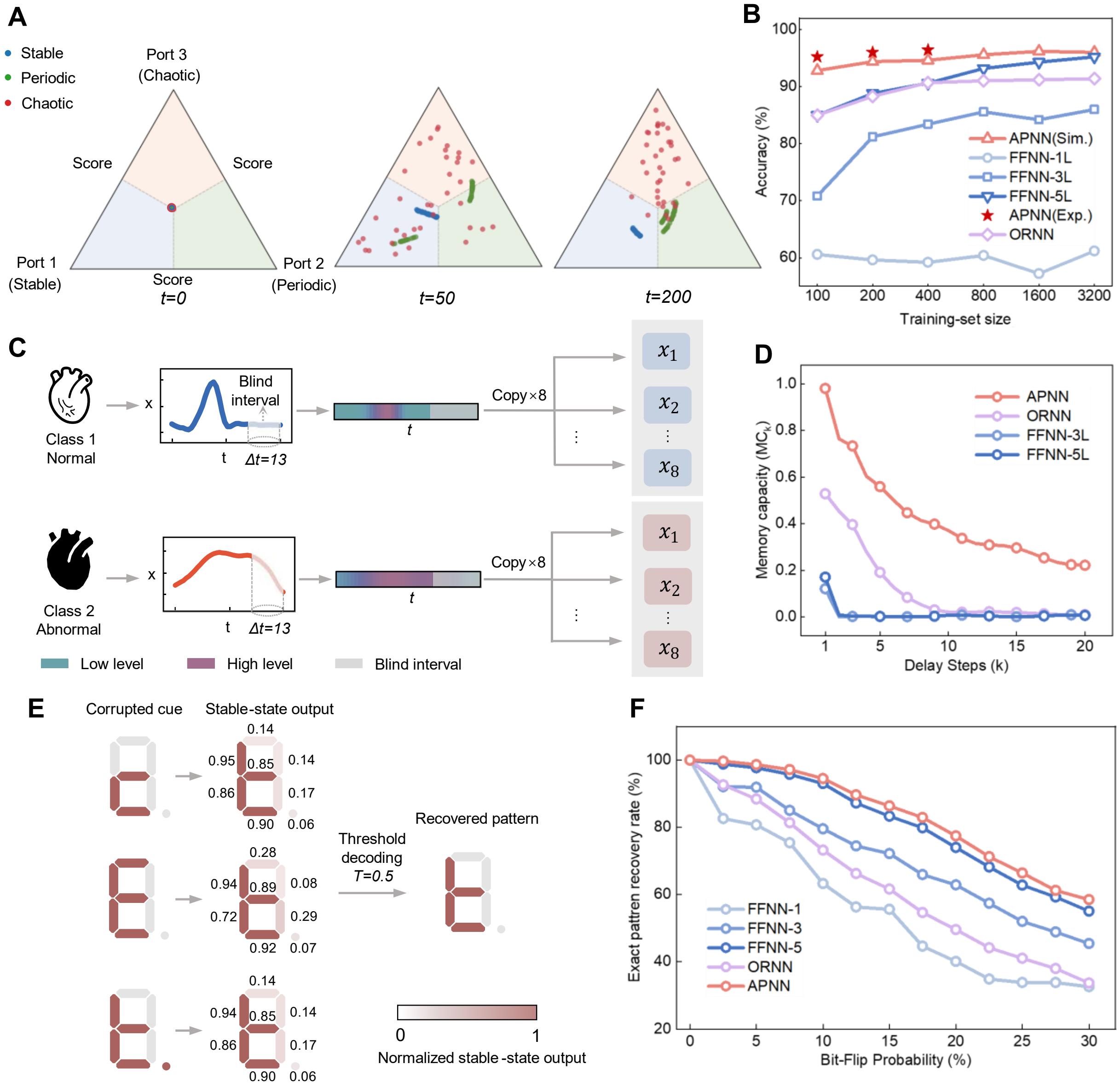

**Fig. 2. Three computational tasks evaluated with the APNN.** (**A**) Simulated responses of the trained network at $G = 2.13$ after $t = 0$, $50$, and $200$ updates. Each point represents one sample; blue, green, and red indicate the three input classes. Ternary coordinates give the normalized contributions of the three output ports, with each vertex representing one port alone. (**B**) Few-shot classification accuracy versus training-set size for the simulated and experimental APNN, a dissipation-dominated optical recurrent neural network (ORNN), and gain-compensated feedforward neural networks (FFNNs) with one layer (FFNN-1L), three layers (FFNN-3L), and five layers (FFNN-5L). (**C**) Experimental workflow for electrocardiogram (ECG) classification using 40 input steps per heartbeat. (**D**) Simulated memory capacity $MC_k$ versus delay steps for the APNN, ORNN, FFNN-3L

and FFNN-5L under ECG sequences. (**E**) Experimental recovery of representative target patterns t from corrupted cues. Color intensity and values show normalized stable APNN outputs; a fixed threshold $T = 0.5$ yields binary recovered patterns. (**F**) Simulated exact eight-bit pattern-recovery rate versus input bit-flip probability for the APNN, ORNN, and FFNN-1L, FFNN-3L and FFNN-5L.

## Nonlinear classification

We used a laser-state classification task in which the classes overlap strongly in the original input space. Simulations and experiments verified that tuning the pump placed the APNN in a near-critical operating regime. Fig. 2A shows the evolution of the trained network at the near-critical operating point: initially overlapping responses to samples from different classes gradually separate over successive recurrent updates. With 100 training samples, the APNN reached 95.2% test accuracy using 44 trainable parameters. A dissipation-dominated ORNN and gain-compensated FFNNs degraded more strongly as the training set was reduced (Fig. 2B). Under the strongest input noise tested, the APNN output mean squared error (MSE) was nearly two orders of magnitude lower than the baselines. These results show that near-critical operation improves the separation and robustness of overlapping inputs with limited training data and parameters.

## Temporal feature extraction

We next classified normal and abnormal ECG heartbeats represented by 40 sequential input steps. The final 13 steps were masked, so the readout had to use information retained from earlier inputs (Fig. 2C). Memory capacity, which quantifies the reconstructibility of a delayed input from the current state, remained approximately 0.22 at a delay of 20 steps in simulation. The ORNN approached zero beyond about 10 steps, and FFNNs retained little information (Fig. 2D). Under the blind-interval protocol, the APNN correctly classified all 60 test heartbeats. These results support the proposed approach to temporal processing: combining finite-bandwidth responses with active recurrent feedback to retain input history and form useful temporal representations within the same physical loop.

## Associative memory

Finally, we trained the recurrent coupling to recover eight-bit target patterns from corrupted cues (Fig. 2E). The stable output was decoded with a fixed threshold, without a separately trained electronic denoiser. The APNN retained a higher exact pattern-recovery rate than the baselines over the tested bit-flip probability (BFP) range (Fig. 2F). At BFP = 30%, its exact recovery rate remained close to 60%, compared with 34% for the ORNN. This shows that the APNN can recover target patterns from corrupted cues under the tested conditions. This supports the proposed use of trained recurrent coupling and driven-dissipative relaxation to form stable, task-consistent outputs, extending the role of the physical loop from discriminative representation to associative recovery.

## Conclusions

We demonstrate a monolithically integrated APNN that harnesses driven-dissipative criticality to address the tension between dynamical sensitivity and stability. Tunable optical driving, finite-bandwidth responses and trained symmetric coupling enable enhanced input sensitivity and extended temporal retention to coexist with stable, task-relevant outputs. On-chip experiments and simulations support discriminative representation, temporal memory and associative recovery within the same recurrent architecture. These results establish controlled driven-dissipative dynamics near criticality as a shared computational resource, offering a new route to efficient and stable physical neural networks.